\documentclass[pdflatex,sn-vancouver-num]{sn-jnl}

\usepackage{graphicx}%
\usepackage{multirow}%
\usepackage{amsmath,amssymb,amsfonts}%
\usepackage{amsthm}%
\usepackage{mathrsfs}%
\usepackage[title]{appendix}%
\usepackage{xcolor}%
\usepackage{textcomp}%
\usepackage{manyfoot}%
\usepackage{booktabs}%
\usepackage{algorithm}%
\usepackage{algorithmicx}%
\usepackage{algpseudocode}%
\usepackage{listings}%
\usepackage{comment} 
\usepackage{graphicx} 
\usepackage{multirow} 
\usepackage{tabularx} 
\usepackage{textcomp} 
\usepackage{array} 
\usepackage{xurl}

\usepackage{ulem}

\theoremstyle{thmstyleone}%
\theoremstyle{thmstyletwo}%

\theoremstyle{thmstylethree}%

\begin{document}

\title[Zero Day Attacks: Novel Behaviour or Novel Vulnerability?]{Zero Day Attacks: Novel Behaviour or Novel Vulnerability?}

\author{\fnm{Nnamdi} \sur{Jibunoh}}\email{njibunoh@nyit.edu}

\author*{\fnm{Sara} \sur{Khanchi}}\email{skhanchi@nyit.edu}

\author{\fnm{Adetokunbo} \sur{Makanju}}\email{amakanju@nyit.edu}

\affil{\orgdiv{Department of Computer Science}, \orgname{New York Institude of technology}, \orgaddress{\city{Vancouver}, \country{Canada}}}


\maketitle


\begin{abstract}
\noindent Zero-day attacks pose severe cybersecurity risks due to their high success rates and stealth. Because signature-based approaches struggle to detect such attacks, building Intrusion Detection Systems (IDSs) for detecting zero-day attacks is essential. We contend that for an IDS to be effective it must be grounded in an understanding of how zero-day attacks manifest in real-world networks. To this end, we review documented zero-day incidents spanning 20 years, finding that these attacks arise from the exploitation of undisclosed vulnerabilities rather than novel attack behavior. Guided by this insight, we propose a taxonomy of zero-day vulnerability types and analyze assumptions of ML-based intrusion detection approaches. Our analysis shows that incidents consistently involve vulnerability exploitation, with memory-corruption flaws being most used; additionally, attacks targeting defensive-mechanism vulnerabilities have increased in recent years. We also identify a mismatch: while incident reports emphasize vulnerability exploitation, many ML-based detectors are designed to detect hypothetical “novel behaviors” during attack execution. Our findings indicate that vulnerability-centric methods are more aligned with real-world attack mechanisms. Consequently, reliance on behavior-based detection alone may overstate zero-day detection capabilities in ML-based IDSs. We advocate for cautious interpretation of such claims and call for improved automated vulnerability detection frameworks aligned with real-world exploit characteristics. Effective defense against zero-day attacks requires prioritizing vulnerability-centeric approaches that enable early identification and mitigation across the lifecycle. The ability to detect attacks that utilize novel behaviors (Tactics, Techniques, and Procedures (TTPs)) is useful, but it does necessarily equate to the ability to detect zero-day attacks.
\end{abstract}




\maketitle

\section{Introduction}
Intrusion Detection Systems (IDSs) remain a critical component of modern defensive security architectures. They are commonly deployed in conjunction with complementary security mechanisms, such as Security Information and Event Management (SIEM) platforms, firewalls, and endpoint protection systems, to provide a layered defense against known and emerging cyber threats \cite{verkerken2024}. This multi-layered model enhances network integrity by ensuring that diverse attack vectors can be rapidly detected, analyzed, and mitigated.

Despite the substantial benefits of IDSs and related security platforms, zero-day attacks continue to pose a major challenge for defenders. According to NIST, zero-day attacks target previously unknown software, hardware, or system vulnerabilities \cite{nist_zero}. Because the underlying vulnerability has never been disclosed or patched, exploitation leaves defenders with “zero days” to respond. One severe aspect of this attack is that the lack of prior signatures or indicators of compromise fundamentally limits traditional detection strategies \cite{guo2023, bhatt2024}. A strong empirical study demonstrates that zero-day vulnerabilities remain undetected for an average of 312 days before disclosure, after which the frequency of exploitation increases by nearly an order of magnitude \cite{bilge2012}.
Similar concerns have been echoed by industry reports such as IBM’s security analysis \cite{ibm_zero} and the Google Cloud Threat Intelligence (CTIG) report of 2024, which identified at least 75 zero-day vulnerabilities disclosed in that year alone \cite{google2024}. Collectively, these findings reinforce the urgent need for improved defensive strategies to detect and mitigate zero-day threats.
`
Over the past decade, Machine Learning (ML)-based IDSs have gained significant research attention because they promise capabilities that exceed those of traditional signature-based systems. ML-based IDSs aim to detect both anomalous system behaviors and known attack signatures, thus offering potential pathways to identify zero-day activity \cite{utwente2021}. A key motivation behind these systems is the possibility of transforming conventional IDSs into adaptive systems capable of evolving as new threats emerge. If achieved, this capability would represent a major advancement in automated cyber defense.

This paper conducts a comprehensive analysis of multi-decade publicly documented zero-day incidents to understand how such attacks manifest in real-world environments. We further propose a taxonomy for categorizing zero-day attacks based on empirical evidence from open-source intelligence (OSINT) datasets spanning the past 20 years. The most critical findings of our analysis include:

\begin{itemize}
  \item Documented zero-day incidents in the real-world all involve the exploitation of unknown/undisclosed hardware/software vulnerabilities
  \item Some of the most extensive publicly-available zero-day attack repositories mainly record only zero-day attacks that involve software vulnerabilities linked to enterprise software. Attacks that involve hardware vulnerabilities, known to exist, are not included. This may be due to the fact that attacks that target software vulnerabilities are more prevalent.
  \item The dominant group of vulnerabilities responsible for zero-day attacks are memory corruption vulnerabilities  like buffer overflow, use-after-free etc.
  \item A recent trend shows that attackers are now targeting defensive mechanism vulnerabilities when carrying out zero-day attacks.
  \item There is a significant misalignment between the operational definition of zero-day attacks used in industry and the definitions implicitly assumed by many ML-based IDS research studies
\end{itemize}

Our last finding is the most significant and we therefore highlight the risks of overstating zero-day detection capabilities in ML-based IDSs that train exclusively on ground-truth datasets derived from known vulnerabilities. Our findings suggest that a shift in focus—from ML-based attack detection to ML-based automated vulnerability assessment—may provide more practical value for mitigating zero-day threats.

The paper is organized as follows. Section~\ref{sec:relatedworks} reviews related work covering ML-based detection, automated vulnerability analysis, and empirical zero-day attack characterizations. Section~\ref{sec:methodology} describes our methodology, including data collection, preprocessing, and taxonomy development. Section~\ref{sec:results} presents the results derived from the analysis of zero-day attacks. Section~\ref{sec:discussion} provides a discussion of key insights, including limitations and implications for zero-day detection. Section~\ref{sec:conclusion} concludes the article with a summary of contributions and recommendations for future research.

\section{Related Work}\label{sec:relatedworks}

This section presents the results of our survey of existing research articles that discuss zero-day attacks. These articles include those that discuss attack detection using ML–based intrusion detection systems and approaches to vulnerability detection. We also examined works that introduce zero-day attack categorization schemes.

\subsection{ML-Based Intrusion Detection Systems (IDSs) for Zero-Day Attack Detection} \label{sec:IDS}

Verkerken et al.\ \cite{verkerken2024} focused on improving the scalability of ML-based IDS solutions by deploying detection components in containerized environments, thus increasing throughput and operational flexibility. Although their work advanced deployment efficiency, the underlying ML models were trained on pre-labeled datasets, which assumes implicitly that previously observed attack classes can generalize to zero-day scenarios. This assumption highlights a critical gap between anomaly detection and genuine novel vulnerability detection.

 Similarly, research on IoT-based IDS for zero-day DDoS attacks, \cite{ddos2021} proposed a honeypot-driven framework to collect attack traffic from Internet-of-Things devices. The data collected was used to train classifiers capable of flagging anomalous traffic patterns. While the system demonstrates improved resilience against previously unseen payload variations, the underlying labeling was derived from a single ground-truth dataset, representing a narrow definition of zero-day events. This introduces bias, as the system effectively learns payload deviations rather than true unknown vulnerability signatures. 

 AlMahmeed and Al-Omay \cite{yusuf2023} introduced the Threat Hunting Intelligence (THI) framework, an iterative threat hunting methodology that integrates operational, tactical and technical intelligence to identify anomalous behaviors from security telemetry sources, including logs and Security Information and Event Management (SIEM) systems. While their multi-pronged detection strategy is operationally relevant, their definition of “zero-day” remains anchored to anomalous activity detection, not novel vulnerability discovery. 

 In contrast, collaborative learning frameworks \cite{collaborative2024} have explored chaining multiple ML-based IDS instances to create a shared knowledge base that can be updated in near real time. Their approach involves training deep feed-forward neural network classifiers on labeled data from newly observed zero-day attacks. However, this still assumes the availability of ground-truth-labeled data prior to model deployment, which conflicts with the very nature of zero-day vulnerabilities, where ground-truth labels are typically unavailable at the time of attack. 

 Yang \cite{guo2023} presented one of the most detailed studies on ML-based zero-day detection, utilizing One-Class SVM and autoencoder-based models to identify anomalous behaviors deviating from a trained “decision boundary".  Although their results reported high zero-day detection rates, the study relied on simulated zero-day data, limiting its ability to evaluate the true performance of the model against unknown vulnerabilities, and this reliance on simulation rather than real-world zero-day exploits underscores a gap between theoretical model performance and operational effectiveness.

In summary, several ML-based Intrusion Detection Systems (IDSs) claim to detect zero-day attacks, yet what they often capture is anomalous behavior rather than attack behavior exploited by a novel zero-day vulnerability. According to NIST, a zero-day vulnerability is an undiscovered or unpatched weakness, which may not lead to a novel pattern of exploitation. Therefore, anomaly detection based IDSs are by definition not trained to detect zero-day attacks. Moreover, reliance on static ground-truth datasets for training, especially those tied to specific vulnerabilities, can skew the generalization capacity of a model and limit its ability to identify truly novel threat vectors \cite{millar2022mlzeroday}. 
Similarly, Montuno et al. \cite{montuno2019ml} highlighted ML’s role in identifying phishing and malware campaigns but cautioned that the effectiveness of such systems is highly dependent on the quality of the training data. Specifically, if an adversary contributes to or influences the ground-truth dataset, they can potentially bias the resulting model, thereby reducing its reliability in operational settings.

These works demonstrate progress in IDS research, but also reveal a consistent methodological assumption: zero-day attacks are assumed to create novel attack patterns, which can be detected by anomaly detection IDSs. However, based on NIST's definition of a zero-day attack, the novelty arises from the unknown vulnerability rather than the attack pattern. This semantic misalignment underscores the need for a refined taxonomy and evaluation framework that more accurately reflects the operational realities of zero-day attack detection.

\subsection{Automatic Vulnerability Detection}
The increasing sophistication and interconnectedness of software systems have made network environments more dynamic and, consequently, more challenging to defend. Traditional intrusion detection and firewall systems have historically been designed to identify known attack signatures; however, their defensive posture alone is often insufficient against evolving threats. As a result, machine learning (ML) has gained traction as a proactive tool for vulnerability detection, particularly in identifying patterns that can indicate the exploitation of unknown or emerging weaknesses \cite{automaticvd2024}.

Abubakar and Pranggono \cite{abubakar2017machine} demonstrated the utility of ML in detecting vulnerabilities within Software-Defined Networks (SDNs), proposing a specialized model capable of identifying denial-of-service attacks. Their findings underscore how automated vulnerability detection can mitigate the high-impact consequences of successful SDN intrusions. 

Google has progressively operationalized ML for vulnerability workflows, moving from ML-based vulnerability triage and prioritization to LLM-assisted discovery and remediation. 
Recent efforts include LLM-generated fuzz targets integrated with OSS-Fuzz and agentic LLM-based vulnerability research (e.g. Big Sleep), alongside LLM-driven pipelines that propose candidate fixes for sanitizer-found bugs for human review \cite{wright2019prioritization,liu2024scaling,ossfuzzgen,bigsleep2024,nowakowski2024patching}.

More recently, Anthropic has demonstrated substantial progress with its Mythos model, which has reportedly identified zero-day vulnerabilities in major operating systems, browsers, and software tools \cite{anthropic2026mythos}. This result highlights the potential of ML-driven approaches to make meaningful progress in vulnerability discovery and zero-day attack prevention.

Mitigating these limitations requires a more robust methodology, including the integration of diverse and continuously updated data sources, the application of adversarial training techniques, and human-in-the-loop validation to contextualize model outputs. Such hybrid approaches can help ensure that ML systems do not conflate payload variability with genuine zero-day discovery. 
\\
\subsection{Categorization of Real-World Zero-Day Attacks}
Several prior studies have sought to classify zero-day incidents, but these efforts differ substantially in scope and granularity. In recent work, an approach centered on \textit{delivery vectors} was proposed that partitions zero-day incidents by their method of delivery, such as malware-based propagation, as well as related factors including configuration weaknesses and software vulnerabilities \cite{delivery2024}. In contrast, our work focuses on the underlying vulnerability characteristics that enable these attacks, emphasizing the root causes of exploitation rather than the external mechanisms through which attacks are delivered.
Bilge and Dumitra\c{s} conducted an influential empirical study of zero-day exploitation in the wild and operationalized categorization using antivirus signature updates, a perspective that emphasizes malware-centric manifestations of previously unknown threats \cite{bilge2012}. These contributions provide valuable operational insights, but do not converge on a unified, delivery-based taxonomy suitable for broader vulnerability-centric analysis. 

 Overall, the literature reveals a methodological flaw: many studies focus primarily on detection performance while insufficiently interrogating the underlying root causes and provenance of the incidents they label as zero-days \cite{guo2023}. This omission has practical consequences: classification schemes derived from detection datasets (for example, signature-updates or payload-based labels) may inadvertently conflate exploitation patterns with distinct vulnerability classes, complicating cross-study comparisons and evaluation of true zero-day discovery capabilities. 

 To address this gap, our work proposes a taxonomy that explicitly separates \textit{observable delivery characteristics} (e.g., payload obfuscation, delivery channel, exploit orchestration) from \textit{vulnerability root causes} (e.g., memory-corruption, logic flaws, hardware weaknesses).  
Prior delivery and malware-oriented taxonomies provide essential empirical detail, but stop short of a systematic, vulnerability-centric classification suitable for validating zero-day discovery. Our contribution therefore complements and extends earlier studies by (i) formalizing a combined delivery/root-cause taxonomy, (ii) providing an annotated dataset of recent incidents consistent with this taxonomy, and (iii) demonstrating how this reclassification changes the interpretation of ML-IDS detection results. 

\noindent 

In summary, existing research demonstrates significant progress in (i) leveraging ML for intrusion detection, (ii) automated vulnerability discovery, and (iii) zero-day attack categorization. However, a closer examination reveals a prevailing reliance on behavioral anomalies or labeled training data as proxies for true zero-day vulnerabilities. This reliance introduces both methodological and operational gaps: ML-based IDSs may excel at identifying novel attack behaviors but fall short of detecting exploitation  patterns due to previously undisclosed vulnerabilities as defined by NIST (zero-day attack pattern). Similarly, most current taxonomies emphasize attack vectors or malware families without adequately capturing the root causes of the underlying vulnerability. Our work addresses this gap by reviewing the definition of zero-day attacks and mapping that to known zero-day attack incidents, introducing a structured delivery and vulnerability-centric taxonomy. This synthesis hopefully provides a more rigorous foundation for evaluating ML-IDS claims and supports more accurate benchmarking of zero-day detection capabilities in operational environments. 

\section{Methodology}\label{sec:methodology}

\begin{figure}[h]
    \centering
    \includegraphics[width=0.85\columnwidth]{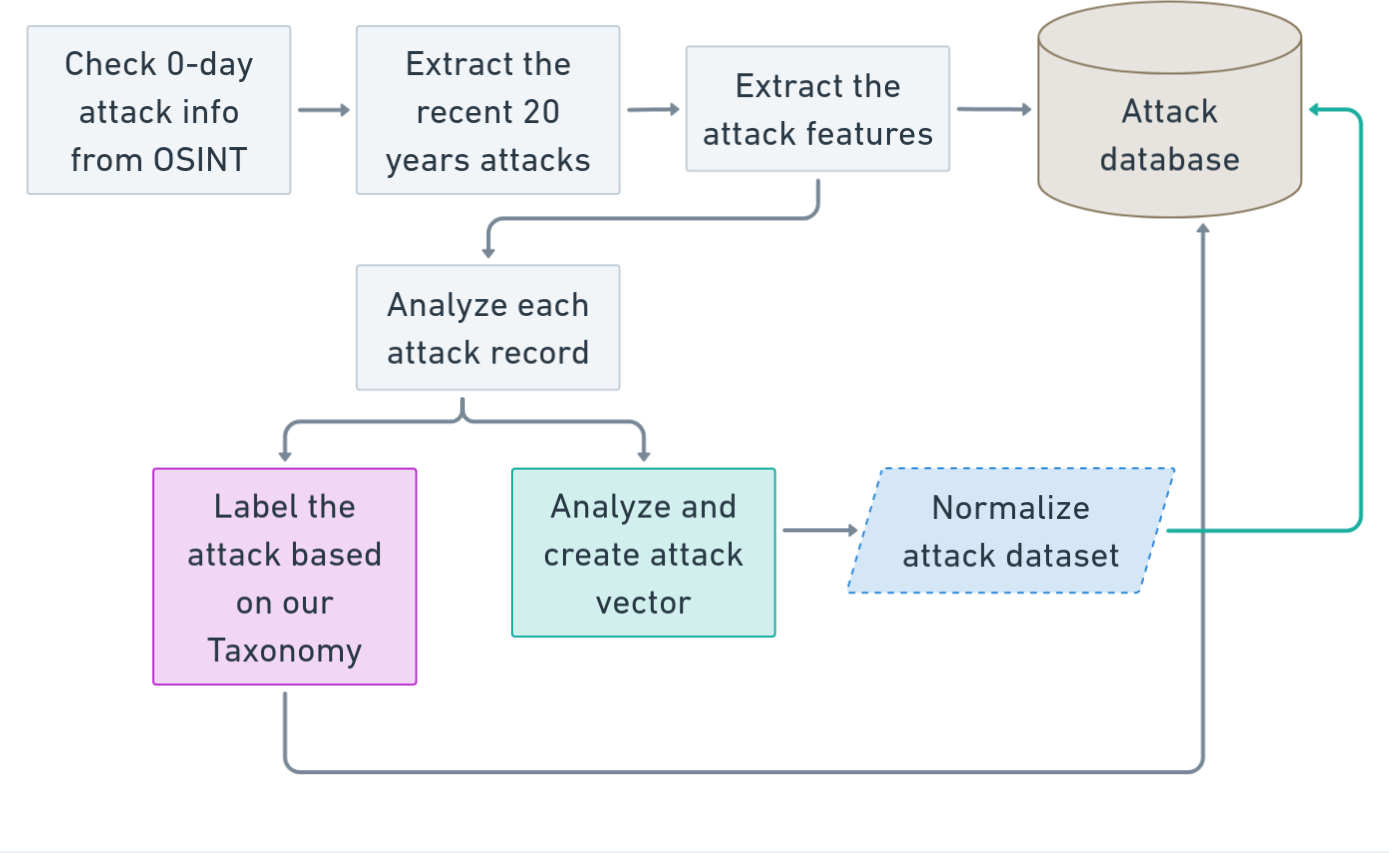}
    \caption{Steps involved in processing all collected zero-day vulnerability data.}
    \label{fig:methodology_flow}
\end{figure}

In this section, we describe the methodology used to collect our zero-day attack data and the process used to develop the zero-day attack taxonomy. The general workflow is illustrated in Figure~\ref{fig:methodology_flow}, following the explanation of each step in the subsections.

We collected real-world zero-day attack information from several OSINT sources, including the National Vulnerability Database~\cite{nist_nvd}, CVE.org~\cite{cve_org}, Palo Alto Unit 42~\cite{unit42-ml-detection}, PortSwigger~\cite{portswigger-nooie}, Google Cloud Threat Intelligence~\cite{google2024} and Google Project Zero~\cite{gpz-zero-day-list}. After analysis, the collected data was reduced to 42 following the methodology shown in Figure~\ref{fig:methodology_flow} and we labeled the dataset Independently Sourced Dataset (INS). This was done in the first phase of our analysis and was instrumental in helping us to figure out the nature of zero-day attacks. Our focus was on any event that could reasonably be classified as a zero-day attack, irrespective of the mechanism of exploitation.  We then reviewed additional repositories that contain historical records of zero-day attacks over the past twenty years. 

For each incident, we extracted relevant features, such as the Common Vulnerabilities and Exposures (CVE) identifier, attack description, root cause, and attack vector. These features provided deeper insight into how each zero-day attack occurred and what the resulting impact was. This enriched dataset formed the basis of our taxonomy.

\subsection{Zero-Day Attack Taxonomy}\label{sec:taxonomy}
This research conducts an in-depth analysis of the nature of zero-day attacks. Using data extracted from documented zero-day incidents, a comprehensive taxonomy is developed to capture and classify the different forms of zero-day attacks.

\subsection{Categorizing Zero-Day Attacks by Novelty}
\begin{figure}[h]
    \centering
    \includegraphics[width=0.95\columnwidth]{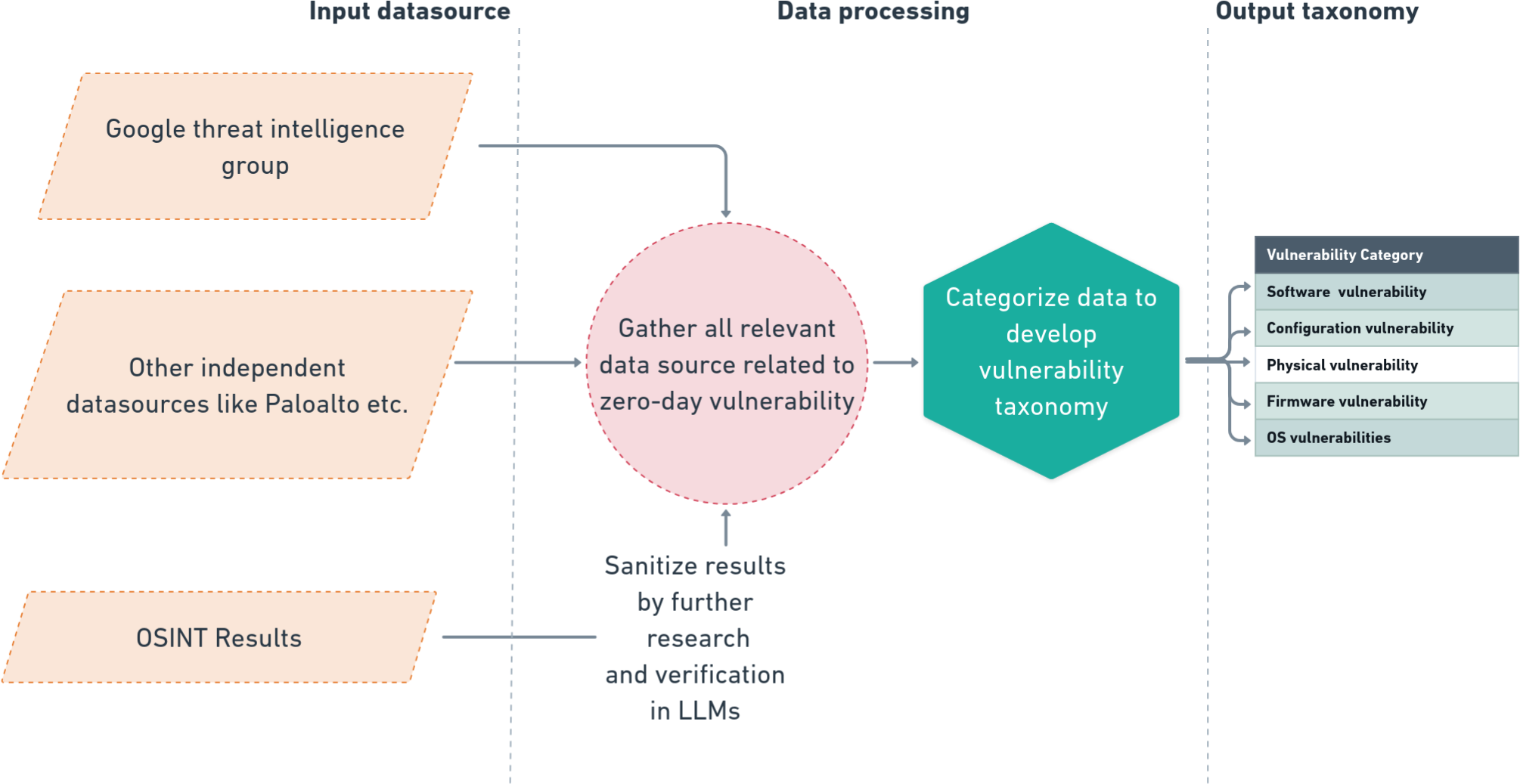}
    \caption{Labeling INS data based on the source of novelty.}
    \label{fig:novelty_flow}
\end{figure}

An essential component of the analysis involved identifying the source of novelty in zero-day incidents and evaluating whether they constitute genuinely novel attacks. Accordingly, the attack vectors from the INS dataset were extracted and systematically analyzed to determine the origin of the novelty (Figure~\ref{fig:novelty_flow}).

\subsection{Zero-Day Attacks of the Past 20 Years}

After coming up with a defined taxonomy based on our initial dataset, INS dataset, we expanded our research to include more zero-day attack incidents to provide a better understanding of the attack history and frequency of zero-day attack types over time.  

In addition to our INS dataset with 42 incidents, we analyzed two independent zero-day attack projects, Google Project Zero (GPZ)~\cite{gpz-zero-day-list} and ZER0-DAY.cz (ZERO-DAY-CZ)~\cite{zerodaycz} with 380 and 892 zero-day records, respectively at the time of the research, providing us with a huge database of attacks to analyze. More incidents are continually being added to these databases.

Both projects classify zero-day attacks based on in-the-wild exploitation of software vulnerabilities prior to public disclosure.

GPZ - founded in 2014 - documents real-world zero-day exploits affecting widely deployed hardware and software systems. Beginning in 2019, GPZ expanded its tracking program to systematically capture zero-day exploits used in the wild. Their dataset includes fields such as CVE, vendor, product, vulnerability type (mapped to Common Weakness Enumeration
(CWE)), description, discovery date, patch date, advisory links, and technical analysis.

ZERO-DAY-CZ - maintained by Cybersecurity Help company, catalogs publicly known zero-day vulnerabilities observed before patches were available. Their dataset includes CVE, title, description, discovery date, patch date, affected software, and CWE classification.

\subsection{Zero-Day CVE Commonality Analysis}
\begin{figure}[h]
    \centering
    \includegraphics[width=0.65\columnwidth]{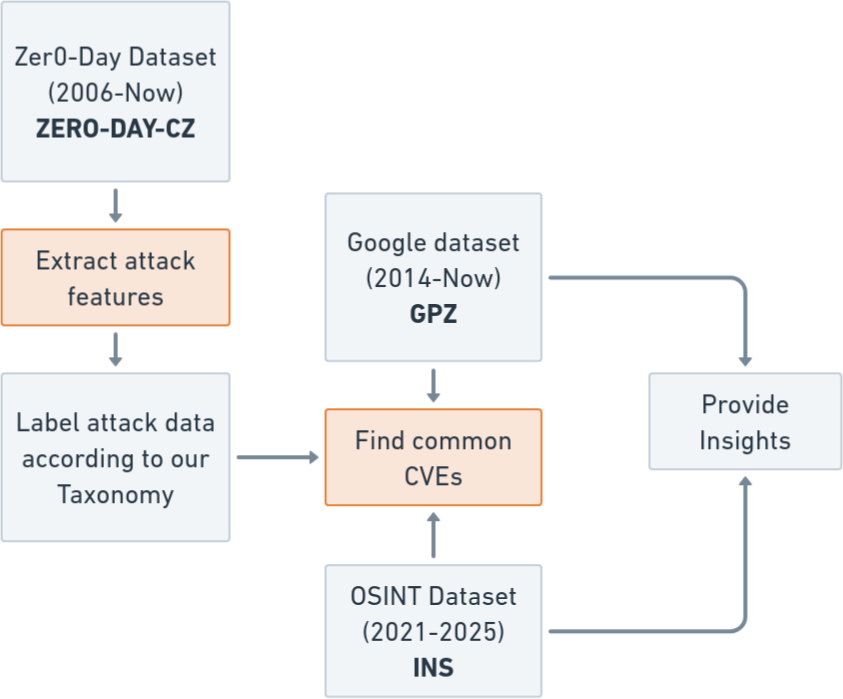}
    \caption{Common vleunerabilities between the three datasets: INS, GPZ and ZERO-DAY-CZ}
    \label{fig:dataset_processing}
\end{figure}

To correlate incidents across datasets, we labeled every record according to our zero-day taxonomy. An initial automated mapping was performed using a structured prompt and ChatGPT, leveraging titles, descriptions, and CWE identifiers of attack incidents. All labels were manually validated to ensure correctness. We then mapped ZERO-DAY-CZ to our INS and GPZ datasets separately based on the CVE number and built a unified view of zero-day trends over the past two decades. This process is highlighted in Figure~\ref{fig:dataset_processing}.

\subsection{Zero-Day Vulnerability Data Availability}
\begin{figure}[h]
    \centering
    \includegraphics[width=0.95\columnwidth]{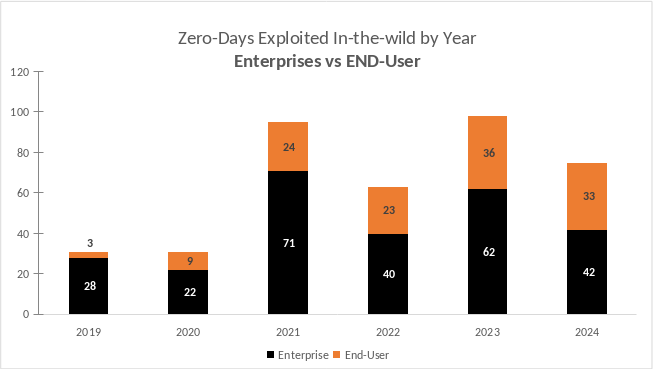}
    \caption{Zero-day exploit Trend: Enterprise vs. END-User}
    \label{fig:dataset_trends}
\end{figure}

We realized early in our research that identifying zero-day vulnerabilities would require additional effort due to the absence of a dedicated zero-day–specific data repository. This challenge motivated the data collection approach described previously under the attack data sources. However, it is important to note that the availability of data  is gradually improving. According to Google’s threat intelligence research~\cite{google2024}, there has been a noticeable increase in the number of independent zero-day vulnerability disclosures from vendors. Figure~\ref{fig:dataset_trends} shows the trend starting in 2021. In the Results section, we discuss the relevance of this increase to our research findings.

\section{Results}\label{sec:results}

In this section, we provide the details of the analysis of our zero-day attack data. The research is carried out using three datasets as described earlier: the INS dataset, which is based on our extensive study of recent zero-day attacks that cover a four year period, and data from the two publicly available zero-day datasets, GPZ and ZERO-DAY-CZ ~\cite{gpz-zero-day-list,zerodaycz}. The results are discussed in detail in the following.

\subsection{Quantitative Analysis}

\subsubsection{Taxonomy of Zero-Day Attacks}
\begin{figure}[h]
    \centering
    \includegraphics[width=0.75\columnwidth]{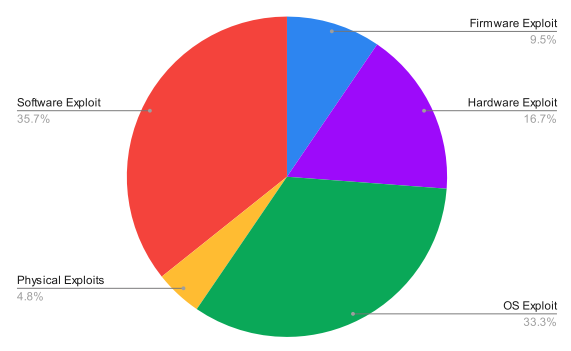}
    \caption{Summary of the identified real-world zero-day attacks for the past four years.}
    \label{fig:four_year_summary}
\end{figure}

Based on the zero-day attack information that we extracted from different zero-day attack incidents (INS dataset), a broad categorization of the incidents in the data can be found in Table~\ref{tab:taxonomy_overview}. This categorization shows that zero-day attacks can be due mainly to software or hardware vulnerabilities, both of which can occur in either design or implementation.\\ 
Based on our analysis of real-world incidents, we derive a detailed taxonomy of attacks categorized by the source of vulnerability, as summarized in Table~\ref{tab:detailed_taxonomy}.
This table illustrates the mapping of the two major categories to our proposed zero-day taxonomy and provides a detailed description of each vulnerability type.
The taxonomy is constructed using the INS dataset, and the corresponding statistics are summarized in Figure~\ref{fig:four_year_summary}.


\begin{table}[h]
\centering
\caption{High-Level Zero-Day Vulnerability Categories}
\label{tab:taxonomy_overview}
\begin{tabular}{|c|p{8cm}|}
\hline
\textbf{Type of Vulnerability} & \textbf{Description} \\
\hline
Hardware Vulnerabilities & Vulnerabilities in microprocessor design, firmware, or peripheral hardware~\cite{cymulate-zeroday}. \\
\hline
Software Vulnerabilities & Software-based vulnerabilities across firmware, operating system, middleware, or application layers. \\
\hline
\end{tabular}
\end{table}

\begin{table*}[t]
\centering
\caption{Detailed Zero-Day Vulnerability Taxonomy}
\label{tab:detailed_taxonomy}

\begin{tabularx}{\columnwidth}{|p{2.5cm}|p{3.4cm}|X|}
\hline
\textbf{Vulnerability Category} & \textbf{Vulnerability Type} & \textbf{Description} \\
\hline

\multirow{5}{*}{\textbf{Software}}
 & Configuration Exploit & Exploitation of insecure or misconfigured systems. \\
 & Software Exploit & Attacks using unknown flaws in application software to perform unauthorized actions. \\
 & OS Exploits & Attacks targeting unknown vulnerabilities in kernels, drivers, or core operating system services. \\
 & Supply Chain Exploit & Zero-day attacks introduced through compromised development, build, or distribution pipelines. \\
 & Firmware Exploit & Exploits targeting firmware components such as UEFI, SMM, or peripheral device controllers. \\
\hline

\multirow{2}{*}{\textbf{Hardware}}
 & Hardware Exploits & Exploitation of micro-architectural or physical design weaknesses. \\
 & Physical Exploits & Attacks requiring direct physical access or device manipulation. \\
\hline

\end{tabularx}

\end{table*}

Our observations of the last four years of zero-day attack data from the INS dataset show that attacks are carried out based on novel vulnerabilities (zero-day vulnerabilities), and not necessarily novel attacks or behaviors after exploitation of the vulnerability. A representative example is the Log4Shell vulnerability (CVE-2021-44228), which impacted a large number of systems. The attack leveraged the JNDI lookup functionality in \texttt{log4j-core}, allowing attacker-controlled input to trigger remote lookups that could result in arbitrary class loading and remote code execution (RCE). While the resulting impact, RCE, is a well-established and widely observed attack outcome, the novelty did not lie in the attack goal itself. Instead, it came from the previously unrecognized vulnerability that enabled RCE through the JNDI feature of a logging framework. In this case, attackers reused a long-known attack objective but achieved it by exploiting an unexpected and newly identified flaw in a commonly trusted component. Thus, the attack technique was not novel; rather, the vulnerability that enabled it in this context was novel.

This vulnerability was present in affected systems for more than eight years before public disclosure. During that period, there was no reported ML-based IDS that detected its presence, although it existed in a widely used library. However, immediately after the public disclosure, multiple vendors were able to detect variations in the initial attack.

As Figure~\ref{fig:four_year_summary} demonstrates, of the total 42 zero-day attacks in the INS dataset, almost 17\% are hardware vulnerabilities (physical and hardware), and the rest are related to some type of software vulnerability. This is close to the common industry definition of zero-day attacks as the exploitation of an unknown software vulnerability.

\subsection{Ancillary Zero-Day Attack Data}

\subsubsection{GPZ Data Analysis}

\begin{figure}[h]
    \centering
    \includegraphics[width=0.85\columnwidth]{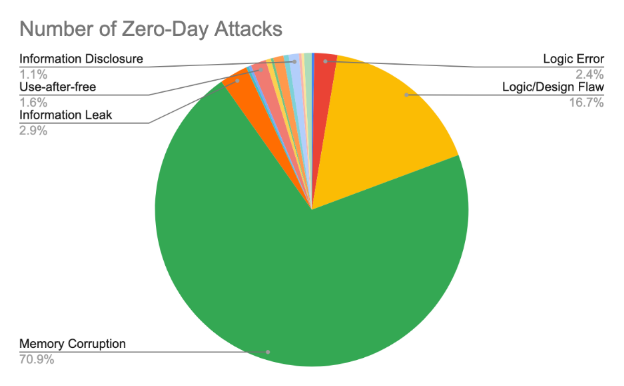}
    \caption{Summary of vulnerability categories in the GPZ zero-day dataset.}
    \label{fig:gpz_summary}
\end{figure}

The total number of zero-day attacks in the wild recorded by GPZ is 380, gathered over the past 11 years (2014--2025)~\cite{gpz-zero-day-list}. All of these zero-day attacks are based on software vulnerabilities. The vulnerability categories are summarized in Figure~\ref{fig:gpz_summary}.

We also calculated a yearly breakdown of the occurrences for each type of vulnerability, summarized in Table~\ref{tab:attack_type_by_year}. The vulnerability type is the label that Google researchers assigned based on CWE values. For this summary, we used the patched date provided in the Google dataset to determine the year column. This choice is motivated by the fact that the field is the most populated date attribute in the dataset and closely approximates the required discovery date. More detailed definitions and descriptions of each category are presented in Table~\ref{tab:category_definitions}.

\begin{table*}[t]
\centering
\caption{Attack type summary breakdown by year from the GPZ dataset.}
\label{tab:attack_type_by_year}
\resizebox{\textwidth}{!}{%
\begin{tabular}{lrrrrrrrrrrrrr}
\toprule
\multirow{2}{*}{\textbf{Type}} & \multicolumn{13}{c}{\textbf{Year}} \\
\cmidrule(lr){2-14}  
 & 2014 & 2015 & 2016 & 2017 & 2018 & 2019 & 2020 & 2021 & 2022 & 2023 & 2024 & 2025 & \textbf{Grand Total} \\
\midrule
Improper authentication       &  &  &  &  &  &  &  &  &  &  & 1 &  & \textbf{1} \\
Information Disclosure        &  &  &  &  &  &  &  & 1 & 1 & 2 &  &  & \textbf{4} \\
Information Leak              &  & 2 & 4 & 2 &  &  & 2 & 1 &  &  &  &  & \textbf{11} \\
Integer overflow              &  &  &  &  &  &  &  &  &  &  & 1 &  & \textbf{1} \\
Logic Error                   &  &  &  &  &  &  &  &  &  &  & 5 & 4 & \textbf{9} \\
Logic/Design Flaw             & 6 & 5 &  & 3 &  & 3 & 3 & 18 & 14 & 11 &  &  & \textbf{63} \\
Memory Corruption             & 5 & 13 & 21 & 16 & 12 & 15 & 20 & 47 & 25 & 42 & 22 & 30 & \textbf{268} \\
PAC bypass                    &  &  &  &  &  &  &  &  &  &  &  & 1 & \textbf{1} \\
Race Condition                &  & 1 &  &  &  &  &  &  &  &  &  &  & \textbf{1} \\
Security Feature Bypass       &  &  &  &  &  &  &  &  &  &  & 3 &  & \textbf{3} \\
Type Confusion                &  & 1 &  &  &  &  &  &  &  &  &  &  & \textbf{1} \\
Unspecified                   &  &  &  &  &  & 1 &  &  &  &  &  &  & \textbf{1} \\
Use-after-free                &  & 6 &  &  &  &  &  &  &  &  &  &  & \textbf{6} \\
UXSS                          &  &  &  & 1 &  & 1 &  &  &  &  &  &  & \textbf{2} \\
XSS                           &  &  &  &  &  &  &  & 1 &  & 1 &  &  & \textbf{2} \\
Other / Unspecified           &  &  &  &  &  &  &  &  & 1 &  &  & 5 & \textbf{6} \\
\midrule
\textbf{Grand Total}          & \textbf{11} & \textbf{28} & \textbf{25} & \textbf{22} & \textbf{12} & \textbf{20} & \textbf{25} & \textbf{69} & \textbf{40} & \textbf{56} & \textbf{32} & \textbf{40} & \textbf{380} \\
\bottomrule
\end{tabular}
}
\end{table*}

\begin{table}[h]
\centering
\caption{Definitions of GPZ dataset vulnerability categories.}
\label{tab:category_definitions}
\renewcommand{\arraystretch}{1.1}
\begin{tabular}{|p{0.3\columnwidth}|p{0.7\columnwidth}|}
\hline
\textbf{Category} & \textbf{Definition} \\
\hline
Improper Authentication & The system fails to properly verify the identity of a user or component, allowing unauthorized access. \\
\hline
Information Disclosure & Sensitive information is exposed to unauthorized users due to inadequate protection or unintended exposure. \\
\hline
Information Leak & Unintended release of internal or confidential data (e.g., through error messages, logs, or side channels); a specific form of information disclosure. \\
\hline
Integer Overflow & An arithmetic operation exceeds the storage capacity of an integer type, causing incorrect calculations or potential security issues. \\
\hline
Logic Error & A flaw in program logic causes it to behave incorrectly or insecurely under certain conditions. \\
\hline
Logic/Design Flaw & A broader category of errors introduced by insecure design decisions rather than coding mistakes (e.g., flawed trust model or protocol). \\
\hline
Memory Corruption & Program memory is modified unexpectedly (e.g., buffer overflow), potentially allowing crashes or arbitrary code execution. \\
\hline
PAC Bypass & A Pointer Authentication Code (PAC) protection mechanism (used in ARM architectures) is bypassed, allowing pointer manipulation and exploitation. \\
\hline
Race Condition & Two or more threads or processes access shared resources in an unexpected order, leading to unpredictable or unsafe behavior. \\
\hline
Security Feature Bypass & Mechanisms intended to enforce security (e.g., ASLR, sandboxing, authentication) are circumvented by an attacker. \\
\hline
Type Confusion & Code incorrectly treats an object as a different type, leading to memory corruption or unauthorized access. \\
\hline
Unspecified & The vulnerability has not been categorized or lacks enough information to assign a specific CWE type. \\
\hline
Use-after-free & The program uses memory after it has been freed, which can result in crashes or arbitrary code execution. \\
\hline
UXSS (Universal Cross-Site Scripting) & A browser-level flaw allowing malicious scripts to bypass same-origin policy and execute in any web domain’s context. \\
\hline
XSS (Cross-Site Scripting) & A web vulnerability that allows attackers to inject and execute malicious scripts in a user’s browser within a specific website’s context. \\
\hline
\end{tabular}
\end{table}

As evidenced by these results, the \textit{memory corruption} vulnerability type is the root cause of almost 71\% of listed CVEs on zero-day attacks and remains persistent over many years. \textit{Logic/design} flaws are ranked second. Earlier data consistently reference \textit{information leaks}; in contrast, recent years have shown a shift toward the use of \textit{information disclosure}, with \textit{logic/design flaws} increasingly categorized as \textit{logic errors}. One of the most notable recent trends is the rise of \textit{security feature bypasses}, suggesting an evolution in attack strategies toward undermining defensive mechanisms rather than exploiting isolated flaws. The remaining categories occur intermittently and do not show persistent trends.\\

\begin{figure}[t]
    \centering
    \includegraphics[width=0.85\columnwidth]{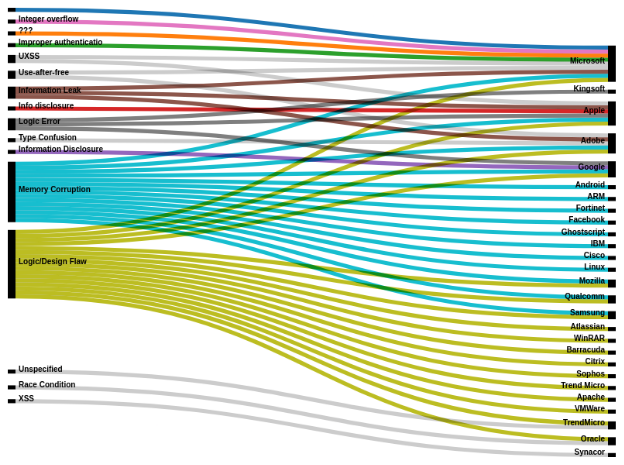}
    \caption{Mapping between attack categories and affected vendors in the GPZ dataset, created using \cite{mauri2017}.}
    \label{fig:vendor_mapping}
\end{figure}

Figure~\ref{fig:vendor_mapping} illustrates the relationship between zero-day vulnerability categories and affected vendors. As represented in this figure and also based on the report of the Google Threat Intelligence Group ~\cite{google2024}, most of the affected systems are enterprise-grade software and security systems. Although the exact reason is not explicitly stated in the data, it is believed that the potential for financial gain and espionage may be driving this behavior.\\

\subsubsection{ZERO-DAY-CZ Data Analysis}
\begin{figure}[h]
    \centering
    \includegraphics[width=0.75\columnwidth]{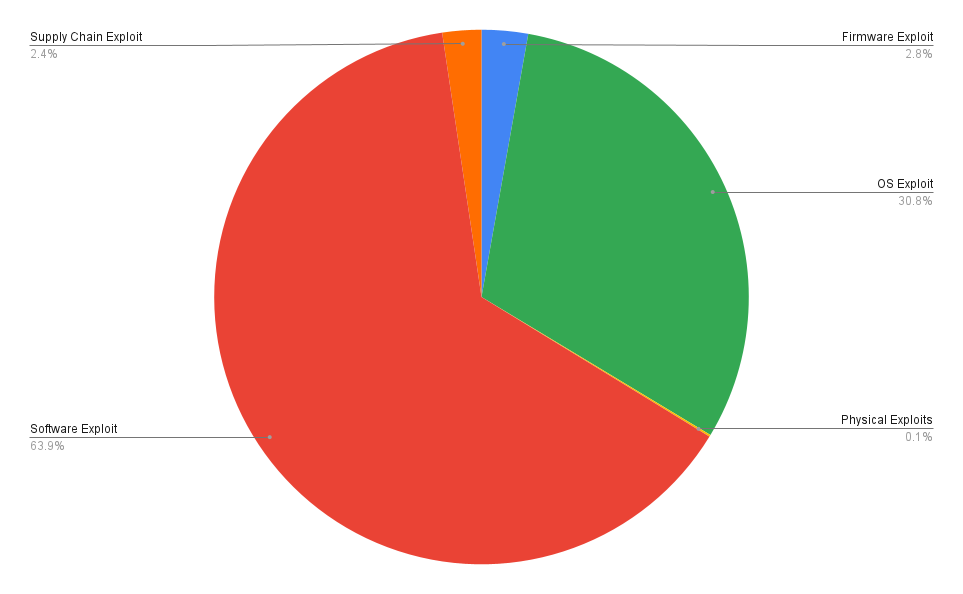}
    \caption{Breakdown of attack types in ZERO-DAY-CZ based on the proposed taxonomy.}
    \label{fig:zerodaycz_taxonomy}
\end{figure}

The total number of records in the ZERO-DAY-CZ dataset is 892, covering the period from 2006 to 2025. Figure~\ref{fig:zerodaycz_taxonomy} illustrates the distribution of attack types according to the taxonomy developed in this study.

\begin{figure}[h]
    \centering
    \includegraphics[width=0.75\columnwidth]{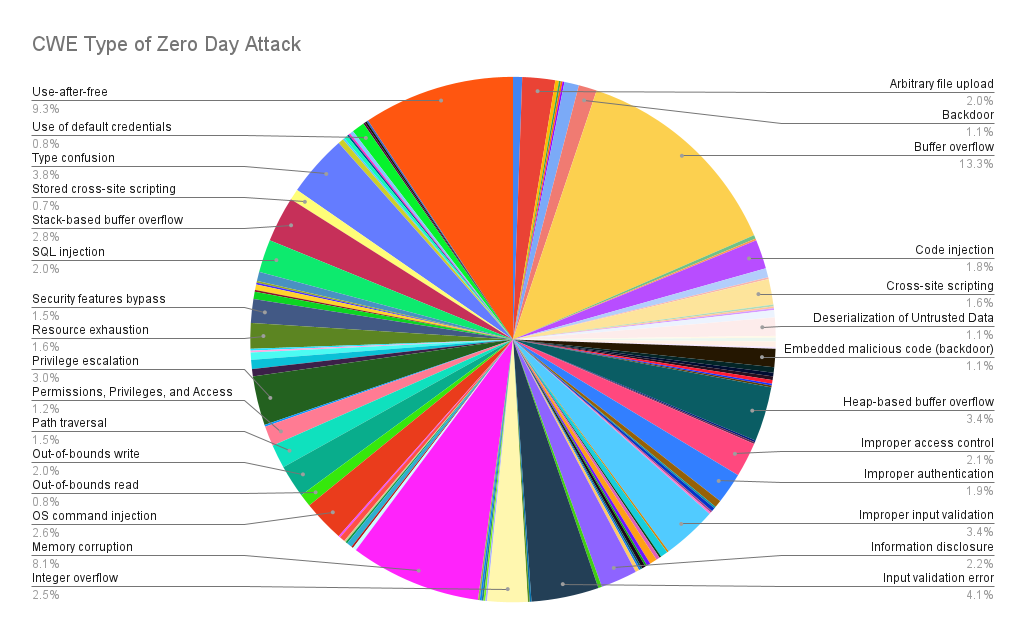}
    \caption{Distribution of CWE types within ZERO-DAY-CZ, highlighting memory corruption-related vulnerabilities.}
    \label{fig:zerodaycz_cwe}
\end{figure}

The distribution of vulnerability types according to the CWE type in the dataset provides further insight (Figure~\ref{fig:zerodaycz_cwe}). This CWE information is more granular than the GPZ categorization; for example, within the memory corruption group, we can identify specific causes such as buffer overflows or out-of-bound writes. Figure~\ref{fig:zerodaycz_cwe} demonstrates that a large proportion of attack types are due to memory corruption vulnerabilities, including different types of buffer overflow, memory corruption issues, and use-after-free bugs.

\subsection{Commonalities Between Datasets}

In this section, we review the common CVEs between the ZERO-DAY-CZ dataset and the GPZ dataset and also the common CVEs between the ZERO-DAY-CZ and INS dataset. We didn't consider the common CVEs between GPZ and INS, as most of the common CVEs are already considered in common CVEs of ZERO-DAY-CZ and GPZ, and only one CVE is left.\\

\subsubsection{ZERO-DAY-CZ and GPZ}

ZERO-DAY-CZ and GPZ overlap for the period from 2014 to the present. Of the 380 CVEs in the ZERO-DAY-CZ project for that period, 354 are also present in GPZ. We compiled these common CVEs into a table and categorized them according to our taxonomy. This approach provides richer content by integrating both GPZ and ZERO-DAY-CZ vulnerability categories alongside our proposed taxonomy. This section presents key insights derived from this merged dataset.\\

Table~\ref{tab:gpz_zeroday_overlap} provides the number of zero-day CVEs in the merged dataset based on the vulnerability type assigned by Google researchers in the GPZ dataset. This summarizes the relative contributions of each vulnerability type to the total number of zero-day attack CVEs.\\

\begin{table}[h]
\centering
\caption{CVE counts by vulnerability type (GPZ \& ZERO-DAY-CZ overlap)}
\label{tab:gpz_zeroday_overlap}
\renewcommand{\arraystretch}{1.1}
\begin{tabular}{|p{4cm}|c|}
\hline
\textbf{Type} & \textbf{Count of CVE} \\
\hline
BLANK & 2 \\
\hline
Not Defined & 4 \\
\hline
Improper Authentication & 1 \\
\hline
Information Disclosure & 4 \\
\hline
Information Leak & 10 \\
\hline
Integer Overflow & 1 \\
\hline
Logic Error & 9 \\
\hline
Logic/Design Flaw & 58 \\
\hline
Memory Corruption & 248 \\
\hline
PAC Bypass & 1 \\
\hline
Race Condition & 1 \\
\hline
Security Feature Bypass & 3 \\
\hline
Type Confusion & 1 \\
\hline
Unspecified & 1 \\
\hline
UXSS & 2 \\
\hline
XSS & 2 \\
\hline
\textbf{Grand Total} & \textbf{354} \\
\hline
\end{tabular}
\end{table}

\begin{figure}[h]
    \centering
    \includegraphics[width=0.95\columnwidth]{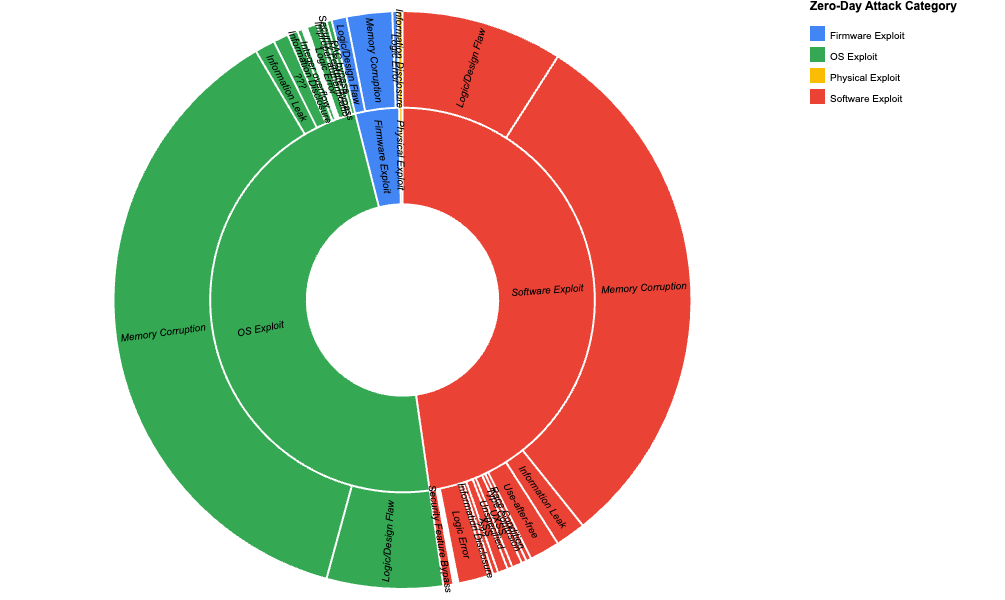}
    \caption{Mapping of GPZ vulnerability types to zero-day attack hierarchy, created using \cite{mauri2017}.}
    \label{fig:gpz_cwe_mapping}
\end{figure}

Figure~\ref{fig:gpz_cwe_mapping} illustrates the distribution of vulnerabilities between GPZ categories within the framework of our proposed taxonomy. As shown, memory corruption constitutes the largest category within each type of exploit. Furthermore, vulnerabilities from different GPZ categories are present across all types in our taxonomy, and it cannot be concluded that any specific GPZ category is exclusively associated with OS, software, or firmware exploit categories.

\begin{figure}[t]
    \centering
    \includegraphics[width=0.75\columnwidth]{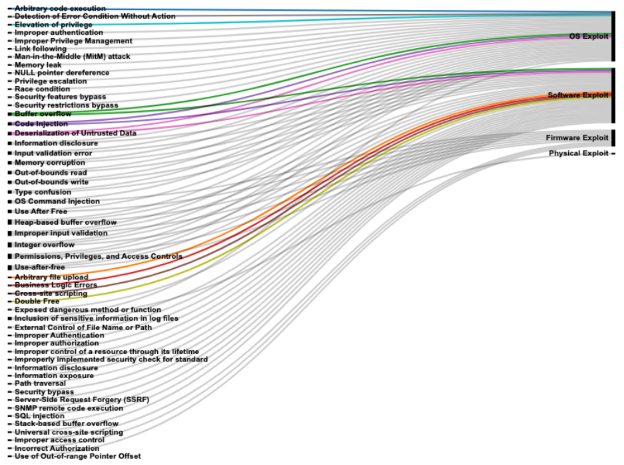}
    \caption{Mapping of CWE categories to the proposed zero-day taxonomy, created using \cite{mauri2017}.}
    \label{fig:cwe_taxonomy_mapping}
\end{figure}

Consequently, Figure~\ref{fig:cwe_taxonomy_mapping} illustrates the mapping of CWEs to our proposed taxonomy. Colored connections highlight CWEs with a strong representation, reflecting a higher number of instances for vulnerabilities such as Buffer Overflow, Code Injection, Deserialization, Arbitrary File Upload, etc. 

Table~\ref{tab:dominant_cwe_summary} summarizes the primary CWE types recorded as vulnerability types in the ZERO-DAY-CZ dataset and maps them to the corresponding GPZ vulnerability categories. This gives a perspective of what CWE types are listed under GPZ categories. \\

\begin{table}[h]
\centering
\caption{Summary of dominant CWE types by GPZ vulnerability type.}
\label{tab:dominant_cwe_summary}
\renewcommand{\arraystretch}{1.1}
\begin{tabular}{|p{4cm}|p{5cm}|c|}
\hline
\textbf{Type (GPZ)} & \textbf{Dominant CWE mappings} & \textbf{Count Summary} \\
\hline
Memory Corruption & Buffer Overflow, Use-after-free, Type Confusion & $>125$ \\
\hline
Improper Input Validation & Improper Input Validation, Input Validation Error, Integer Overflow & 46 \\
\hline
Logic/Design Flaw & Logic Error, Business Logic Errors & 8 \\
\hline
Security Feature Bypass & Security restrictions bypass, permissions/privilege flaws & 10 \\
\hline
Information Disclosure & Information disclosure / exposure / leak & 8 \\
\hline
Type Confusion & Type Confusion (CWE-843) & 29 \\
\hline
Use-after-free & Use-after-free (CWE-416) & 50 \\
\hline
\end{tabular}
\end{table}

\subsubsection{ZERO-DAY-CZ and INS}

\begin{figure}[t]
    \centering
    \includegraphics[width=0.95\columnwidth]{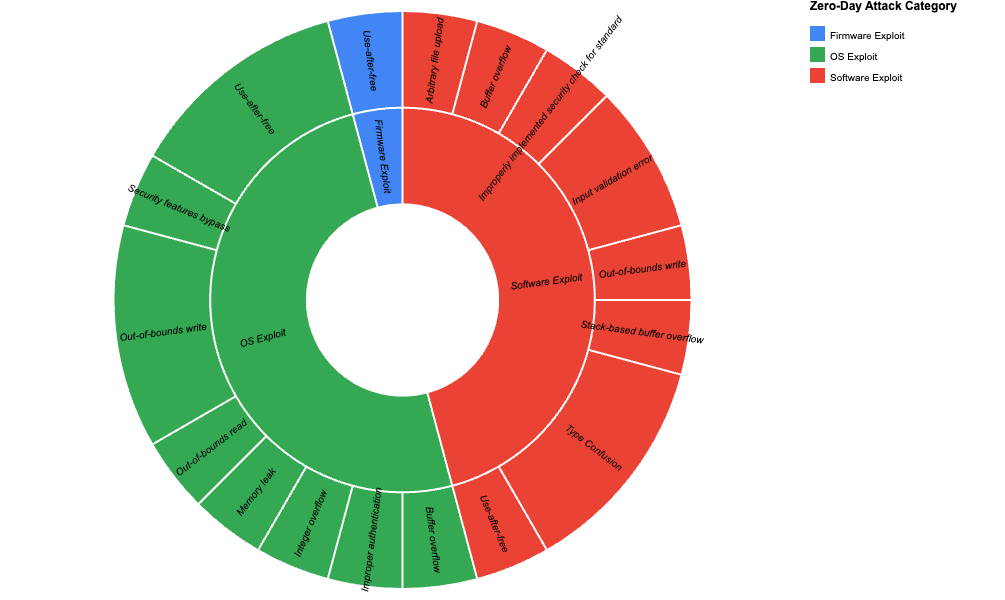}
    \caption{Zero-day attack categories and CWE types for common CVEs between INS and ZERO-DAY-CZ datasets, created using \cite{mauri2017}.}
    \label{fig:ins_zerodaycz_common}
\end{figure}

There are 25 common zero-day attacks between the INS and the ZERO-DAY-CZ datasets. All these common CVEs fall into three main categories: OS, software, and firmware vulnerabilities, in that order of prevalence. Figure~\ref{fig:ins_zerodaycz_common} demonstrates the zero-day attack categories and their respective CWE types for the common CVEs between the two datasets. Here, again, we see the distribution of CWE types in each exploit type in our proposed category. The most common types of vulnerability include use-after-free, type confusion, out-of-bounds errors, and input validation errors.\\

\section{Discussion} \label{sec:discussion}

\subsection{Terms and Definitions}

In this section, we begin by reviewing relevant terms and definitions, starting with what characterizes a zero-day attack. 

A zero-day attack occurs due to the exploitation of an unknown vulnerability. For a vulnerability to be classified as zero-day, it must satisfy the following criteria:

\begin{enumerate}
    \item The vulnerability must be a previously unknown hardware, firmware, or software flaw~\cite{nist_zero}.
    \item There must be evidence that the vulnerability was discovered and exploited before vendor awareness or public disclosure.
\end{enumerate}

Another important concept is the \textit{detection of anomalous behavior}. For an ML-based IDS to detect anomalous behaviors, it must be capable of identifying:

\begin{itemize}
    \item Statistical deviations from learned normal system or network behavior.
    \item Payload variations and obfuscated versions of known attacks.
    \item System or network activity outliers.
    \item Suspicious traffic patterns indicative of exploitation attempts.
\end{itemize}

Following from the definitions above, a \textit{novel attack} technique that exploits a known vulnerability does not qualify as a zero-day attack. A zero-day attack requires the exploitation of a vulnerability that is itself undisclosed or unknown. Misclassifying novel behaviors (e.g., variations of buffer overflow, XSS, etc.) as zero-day leads to inflated detection-performance claims and a misrepresentation of the threat landscape.

Our analysis shows that although zero-day attacks begin with unknown vulnerabilities, most ultimately follow familiar attack patterns. The novelty lies in the vulnerability exploited, not in the attack technique. Real‑world zero‑day attacks including Spectre and Meltdown side‑channel exploits, the Heartbleed semantic memory‑safety flaw, OAuth authorization bypass zero‑days in identity providers, and business‑logic vulnerabilities in cloud services often exhibit statistically normal behavior, exposing a fundamental limitation of anomaly‑based ML intrusion detection as a means of detecting zero-day attacks. Thus, mitigating zero-day attacks requires prioritizing \textit{vulnerability discovery and assessment}, not simply anomaly detection.

In summary, zero-day attacks are defined by exploitation of an unknown vulnerability, giving defenders “zero days” to react. In contrast, novel attack behavior refers to new tactics, techniques, or procedures (TTPs) that exploit known weaknesses. Although both are emerging threats, they are fundamentally different.

\subsection{Research Analysis Framework}

\begin{figure}[h]
    \centering
    \includegraphics[width=0.55\columnwidth]{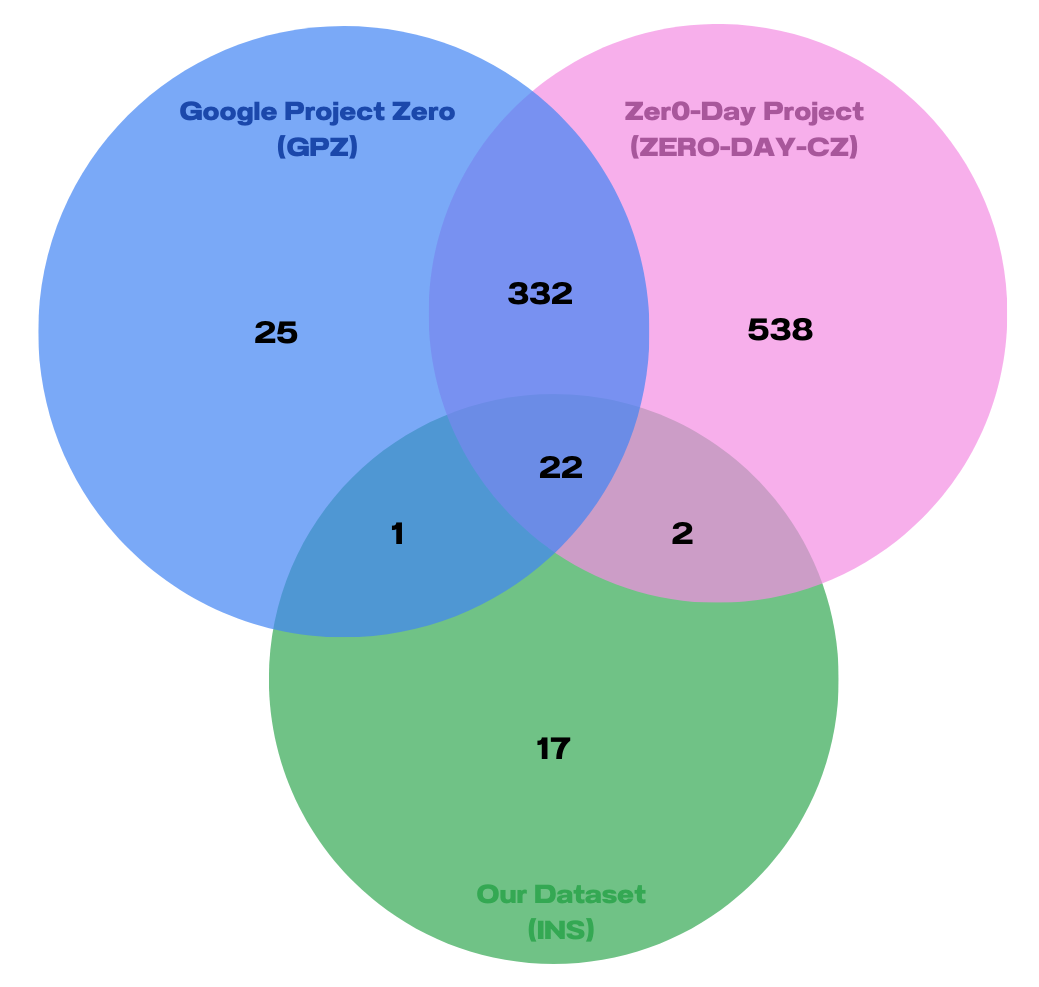}
    \caption{Venn diagram showing intersection of zero-day attack datasets.}
    \label{fig:discussion_fig_12}
\end{figure}

Our working dataset contains 937 unique zero-day vulnerability records from three primary sources: the GPZ dataset (n = 380), the ZERO-DAY-CZ database (n = 892), and our INS dataset (n = 42). Figure~\ref{fig:discussion_fig_12} shows the Venn diagram that illustrates the overlap in these datasets.

Our analysis reveals that GPZ and ZERO-DAY-CZ include only software-based zero-day vulnerabilities. None of the hardware-based vulnerabilities present in our dataset appear in those datasets, despite Google's indication that hardware vulnerabilities are included in their methodological scope.

Across all datasets, Memory Corruption vulnerabilities dominate at approximately 71\%, followed by Logic/Design Flaws at 19\%, consistent with our internal taxonomy.

Further inspection of patch timelines shows:

\begin{itemize}
    \item Mean zero-day window: 34.2 days
    \item Median zero-day window: 9 days
    \item Range: 1–834 days
\end{itemize}

Approximately 47.7\% (63 out of 344) of the vulnerabilities were patched within 7 days of disclosure, while around 25\% remained unpatched for more than 30 days. This suggests that many so-called zero-day “detections” were actually detections of anomalous behavior that occurred \textit{after} disclosure, not genuine zero-day discovery~\cite{bilge2012}.

A similar trend appears in the ZERO-DAY-CZ dataset:  
71.3\% patched the same day, 5.9\% patched within 1–3 days, and 5.8\% within 4–7 days; only 1.9\% took more than 90 days. Compared to findings by Bilge and Dumitra\c{s}~\cite{bilge2012}, who report an average zero-day lifespan of approximately 320 days, these statistics strongly indicate that the majority of detections in these datasets could identify anomalous system behavior rather than discovering unknown vulnerabilities.

Recent advances in ML-based vulnerability discovery further reinforce this interpretation. For example, Anthropic’s Mythos model has demonstrated the ability to identify thousands of previously unknown vulnerabilities across major operating systems, browsers, and widely used software components, including flaws that persisted for decades despite extensive human auditing and automated testing \cite{anthropic2026mythos}. Notably, reported examples include a 27-year-old vulnerability in OpenBSD and a 16-year-old flaw in FFmpeg, illustrating that many exploitable weaknesses remain latent in production systems for extended periods. These findings align closely with our observation that zero-day attacks are fundamentally driven by the exploitation of undisclosed vulnerabilities rather than the emergence of novel attack behaviors. At the same time, they highlight a critical gap in current IDS evaluation practices: while real-world zero-day incidents, and emerging tools such as Mythos, center on vulnerability discovery, many ML-based IDSs continue to focus on detecting anomalous behaviors that often manifest only after disclosure or exploitation. This further supports our conclusion that existing evaluation methodologies may overestimate zero-day detection capabilities and underscores the need to prioritize vulnerability-centric approaches for both detection and system design.

\section{Conclusion} \label{sec:conclusion}

The continued relevance of intrusion detection systems (IDSs) is underscored by the growing demands of an increasingly interconnected world. Recent advances in Large Language Models (LLMs) and machine learning (ML) have further enhanced IDS capabilities, introducing novel heuristic functionalities that complement the reliability and efficiency of traditional signature-based approaches. However, our findings challenge the prevailing assumption that ML-based IDSs are effective at detecting true zero-day attacks. Instead, the evidence suggests that these systems primarily detect variations or manifestations of known attacks, often after vulnerabilities have been disclosed, rather than identifying attacks that were executed via the exploitation of previously unknown vulnerabilities. \\
This limitation reflects a broader mismatch between how zero-day attacks manifest in practice and how ML-based IDSs are typically designed and evaluated. Real-world zero-day incidents are consistently driven by the exploitation of undisclosed vulnerabilities, whereas many ML-based approaches focus on detecting anomalous or “novel” behaviors during attack execution. While such behavioral signals may be indicative of malicious activity, they do not necessarily correspond to the exploitation of the underlying vulnerability that defines a zero-day condition. As a result, current evaluation practices may overstate the extent to which these systems address the core challenge of zero-day detection. \\
These findings highlight the need to reorient research efforts in ML-Based zero-day attack detection toward vulnerability-centric approaches. In particular, greater investment in ML-based tools for vulnerability discovery, especially when integrated into the software and hardware development lifecycle, offers a more promising path for mitigating zero-day risk. Embedding such capabilities within the software development lifecycle (SDLC) can enable earlier identification and remediation of vulnerabilities, leveraging vulnerability-centric guidance such as the OWASP Top 10, while aligning with frameworks like MITRE ATT\&CK to contextualize how such vulnerabilities may be exploited in practice. \\
Beyond technical advances, improved governance and standardization are essential. The taxonomy introduced in this work is a step toward more consistent characterization of zero-day vulnerabilities. Building on this, future work could focus on establishing a centralized and systematically curated registry of zero-day incidents to improve transparency, enable more rigorous evaluation, and support collaborative research. Such efforts would help bridge the gap between theoretical detection approaches and the realities of zero-day exploitation, ultimately contributing to more effective and reliable defensive systems.


\bibliography{refs}

\end{document}